\documentclass[copyright,creativecommons]{eptcs} 

\usepackage{amsmath, amssymb}
\usepackage{tikz}
\usetikzlibrary{patterns, intersections, positioning, shapes.geometric, calc,
automata, arrows}
\usepackage{amsthm}
\usepackage{subcaption}

\newtheorem{definition}{Definition}
\newtheorem{example}{Example}
\newtheorem{theorem}{Theorem}
\usepackage{thmtools}

\newcommand{\play}{\textnormal{play?}}
\newcommand{\quit}{\textnormal{quit?}}
\renewcommand{\repeat}{\textnormal{repeat?}}
\newcommand{\quitRepeat}{\textnormal{quitRepeat?}}
\newcommand{\playList}{\textnormal{endPlayList!}}
\newcommand{\song}{\textnormal{song!}}

\newcommand{\print}{\textnormal{print?}}
\newcommand{\scan}{\textnormal{scan?}}
\newcommand{\printed}{\textnormal{printed!}}
\newcommand{\scanned}{\textnormal{scanned!}}

\usepackage{mathtools}
\newcommand\defis{\;\smash[t]{\stackrel{\mathclap{\normalfont\mbox{\tiny def}}}{=}}\;}

\newcommand{\figureref}[1]{Figure~\ref{#1}}
\newcommand{\exampleref}[1]{Example~\ref{#1}}
\newcommand{\definitionref}[1]{Definition~\ref{#1}}
\newcommand{\theoremref}[1]{Theorem~\ref{#1}}
\newcommand{\sectionref}[1]{Section~\ref{#1}}
\newcommand{\A}{{\cal A}}
\newcommand{\T}{\mathcal{T}}
\newcommand{\pass}{\textit{Pass}}
\newcommand{\fail}{\textit{Fail}}

\newcommand{\afterop}{\mathbin{\textit{after}}}
\newcommand{\after}[2]{{{#1}\afterop{#2}}}
\newcommand{\afterm}[2]{\after{#1}{#2}}
\newcommand{\outop}{\mathop{\textit{out}}}
\newcommand{\out}[1]{\outop({#1})}
\newcommand{\outm}[1]{\out{#1}}
\newcommand{\inpop}{\mathop{\textit{in}}}
\newcommand{\inp}[1]{\inpop({#1})}

\newcommand{\stracesop}{\mathop{\textit{traces}}}
\newcommand{\straces}[1]{\stracesop({#1})}
\newcommand{\stracesm}[1]{\straces{#1}}

\newcommand{\trace}{{\it trace}}
\newcommand{\gtracesop}{\mathop{\textit{trace}}}
\newcommand{\gtraces}{\gtracesop}

\newcommand{\ioco}{\mathrel{\textit{ioco}}}

\newcommand{\Outc}{\textit{Outc}}
\newcommand{\winp}{{\textit{Win$\Pi$}}}
\newcommand{\wins}{{\textit{Win$\Sigma$}}}
\newcommand{\pref}{\textit{Pref}}

\newcommand{\act}{\textrm{Act}}
\newcommand{\reset}{\textit{reset?}}
\newcommand{\moves}{\textit{Moves}}
\renewcommand{\stop}{\textnormal{stop?}}
\newcommand{\stopState}{\bot}
\newcommand{\qgame}{Q_\stopState}
\newcommand{\playpref}{{\Pi^{\it pref}}}

\newcommand{\mydownarrow}{{\downarrow\!}}
\newcommand{\myuparrow}{{\uparrow\!}}
\newcommand{\game}{{G}}

\newcommand{\impl}{{\it Impl}}
\newcommand{\spec}{{\it Spec}}
\newcommand{\B}{{\cal B}}
\newcommand{\alttraceincl}{{\sqsubseteq_2}}

\newcommand{\actions}{{\it actions}}

\title{Tester versus Bug: \\ A Generic Framework for Model-Based Testing via Games}

\date{}

\author{Petra van den Bos $^{1}$ 
  \thanks{Supported by NWO project SUMBAT (grant 13859)} 
  \qquad Marielle Stoelinga $^1$ $^2$ $^*$ 
  \thanks{Supported by NWO project SEQUOIA (grant 15474), NWO project BEAT (grant  612001303), and EU project SUCCESS.}
  \email{\qquad petra@cs.ru.nl \qquad\qquad marielle@cs.utwente.nl}
  \institute{$^1$Institute for Computing and Information Sciences, Radboud University, the Netherlands}
  \institute{$^2$Formal Methods \& Tools, University of Twente, the Netherlands}
}

\def\titlerunning{Tester versus Bug: A Generic Framework for Model-Based Testing via Games}
\def\authorrunning{Petra van den Bos \& Marielle Stoelinga}

\begin{document}

\maketitle
 
\begin{abstract}
 \textbf{Abstract.} We propose a generic game-based approach for test case generation. 
We set up a game between the tester and the System Under Test,
in such a way that test cases correspond to game strategies, and the conformance relation ioco corresponds to alternating refinement.  
We show that different test assumptions from the literature can be easily incorporated, by slightly varying the moves in the games and their outcomes. In this way, our framework allows a wide plethora of game-theoretic techniques to be deployed for model based testing.
 \bigskip
\end{abstract}

\noindent
Testing is a widely practiced method to validate the correctness of a system. By exercising a System Under Test (SUT) to a large number of test cases, the tester gets insight in the quality of the system, and especially whether or not a system conforms to its specification. 
At the same time, testing is very expensive, often requiring 30\% of the development cost of a system. 
Therefore, testing is intrinsically an optimization problem: the tester tries to maximize the effectiveness of the test cases, at a minimum amount of resources. 

This optimization problem is naturally modeled as a two-player game between the tester and the SUT:
in each system state, the tester chooses one of the enabled input actions, and the SUT chooses an enabled output action. Thus, the inputs are under control of the tester, but the outputs of the SUT are not; which closely fits the nature of 2-player games.

Various authors have fruitfully pursued this idea and used game-based approaches to obtain effective testing strategies: this idea was first outlined in \cite{Papadimitriou_2001}, and later refined in several ways, deploying Markov decision processes \cite{VCGST04}, timed games \cite{realtimegames}, and coverage games \cite{nodecoverage}. 
While they have shown their effectiveness in realistic settings, these methods are rather pragmatic in nature. A systematic comparison between model-based testing and the underlying game-theoretic methods is currently missing. 

Indeed, existing game-based testing approaches formulate their methods in terms of a game graph between the tester and the SUT.
Testers, however, do not think of testing as a game graph, but rather in terms of a system specification, from which tests are then derived. 
This means that earlier approaches leave implicit the step from the specification to the  game graph. This step is crucial, since this is where the assumptions on the interaction between the tester and the SUT are encoded. In this paper, we show how different test assumptions are encoded by varying the move outcomes when the tester and the SUT propose their actions. In this way, we obtain a game based testing framework that is generic and overarches the most common testing assumptions.

Further, earlier approaches do not say how game strategies relate to test cases, hence it is not clear if game-based strategies and test cases have the same capabilities. In this paper, we show that each test case corresponds to a finite, trace based game strategy and vice versa, for history-dependent and deterministic strategies that work on a slightly extended state space. Finally, we establish a fundamental connection between testing and games, by showing that 
game refinement corresponds to conformance. More precisely, we show that the so-called Player 2 alternating trace inclusion coincides with the canonical conformance relation in testing, namely
the input-output conformance relation $\ioco$. 

One could argue that current game-based approaches to testing (including ours) focus on the competitive aspect of the interaction between two players: The tester plays \emph{against} the SUT, so that the SUT makes the tester's life as difficult as possible. This means that winning strategies are too pessimistic: in reality, a tester and a SUT have a less hostile relationship. Therefore, an important topic for future work is to relax these assumptions, e.g. by deploying resilient strategies \cite{synthesissafety}.

\paragraph{\bf Contributions.} This paper establishes a fundamental connection between model-based testing and 2-player concurrent games. 
In particular, we derive from a specification automaton a two player game, such that:
\begin{itemize}
\item {\em Test assumptions are explicitly encoded.} This makes our approach generic, so that game theoretic methods can be applied independent of the test assumptions. 
\item {\em Test cases correspond to game strategies} and therefore tests derived via classical test derivation algorithms, and game strategies derived via game theoretic  algorithms have the same power. 
\item {\em Test case derivation corresponds to strategy synthesis}, so that any synthesis algorithm can be deployed to obtain (optimal) test cases. 
\item {\em Conformance coincides with alternating trace inclusion.} This provides a connection between two fundamental preorders. 
\end{itemize}

\paragraph{\bf More related work.} As stated, various approaches to game-based test case deriviation exist. In particular, 
\cite{VCGST04} uses $1\frac{1}{2}$-player games (i.e., Markov decision processes), to obtain cost-optimal test strategies, under the assumptions that the SUT behaves stochastically;
\cite{david2008cooperative,realtimegames,david2009timed} use timed games to find cost-optimal test cases for real-time games. 
In \cite{bloem2016synthesizing}, games are used to generate test cases for Mealy machines based on CLT properties. Since Mealy machines are synchronous, the interaction between the tester and the SUT is much simpler, i.e. the players play in turns.

 In \cite{nodecoverage}, games are used to optimize state coverage, i.e., the number of (specification) states visited during testing; \cite{coveragecomplexity} studies the computational complexity of this problem. 
Finally, \cite{learningIOautomata,alternatingveanes} study the relation between game refinement (more precisely, alternating simulation), and the \emph{ioco} conformance relation. 

\section{Games}

We consider games played  by two players on a game graph.
In each state, both players choose one of their enabled actions, and together these determine the next states the game can be in.
Since both actions are used simultaneously by a {\moves} function to decide on the next state, game arenas (\definitionref{def:gamearena}) describe concurrent 2-player games.

\begin{definition}\label{def:gamearena}
 A \emph{game arena} is a tuple $\game=(Q,q_0,\act_1,\act_2,\Gamma_1,\Gamma_2, \moves)$ where, for $i=1,2$:
 \begin{itemize}
  \item $Q$ is a finite set of states,
  \item $q_0 \in Q$ is the initial state,
  \item $\act_i$ is a finite and non-empty set of Player $i$ actions,
  \item $\Gamma_i: Q\to 2^{\act_i} \setminus \emptyset$
  is an enabling condition, which assigns to each state $q$ a non-empty set $\Gamma_i(q)$ of actions available to Player $i$ in that state, and
  \item $\moves: Q \times \act_1 \times \act_2 \rightarrow 2^Q$ is a function that
 given the actions of Player 1 and 2 determines the set of next states $Q' \subseteq Q$ the game can be in.
 We require that $\moves(q,a,x) = \emptyset$ iff $a\not\in\Gamma_1(q) \vee x\not\in\Gamma_2(q)$.
  \end{itemize}
\end{definition}
A \emph{play} is an infinite path in a game arena, i.e., a sequence of states and actions of both players. 
We consider prefixes of plays as their finite description.
A play is winning if it visits some state in a reachability goal $R \subseteq Q$.

\begin{definition} \label{def:play}
 A \emph{play} of a game arena $\game=(Q,q_0,\act_1,\act_2,\Gamma_1,\Gamma_2, \moves)$ is an infinite sequence:
 \[\pi = q_0\langle a_0,x_0 \rangle q_1 \langle a_1, x_1 \rangle  q_2 \dots \]
 with $a_j\in\Gamma_1(q_j)$, 
 $x_j\in\Gamma_2(q_j)$, and 
$q_{j+1} \in \moves({q_j},\allowbreak{a_j},{x_j})$ for all $j \in \mathbb{N}$.
We write $\pi_j^q \defis q_j$, $\pi_j^a \defis a_j$, and $\pi_j^x \defis x_j$ for the $j$-th state, player 1 action, and player 2 action respectively.
The set of all plays of $\game$ is denoted $\Pi(\game)$.

We define $\pi_{0:j} \defis q_0\langle a_0,x_0 \rangle q_1\allowbreak \langle a_1, x_1 \rangle  q_2 \dots q_j$ as the prefix of play $\pi$ up to the $j$-th state.
With $|\pi|$ we denote the \emph{length} of a prefix $\pi$, i.e. the number of states in $\pi$.
 The set of all prefixes of a set of plays $P \subseteq \Pi(\game)$ of $\game$ is denoted $\pref(P) \defis \{\pi_{0:j} \mid \pi \in P, j \in \mathbb{N}\}$.
 We define $\playpref(\game) \defis \pref(\Pi(\game))$.
 \end{definition}
 
 \begin{definition} \label{def:winning}
 A play $\pi \in \Pi(\game)$ of a game arena $\game$ is {\em winning} with respect to reachability goal $R \subseteq Q$, if $\pi$ reaches some state in $R$, i.e., there exist a $j \in \mathbb{N}$ such that $\pi_j^q \in R$.
 We write $\winp(\game,R)$ for the set of winning plays with respect to $R$. 
\end{definition}

The players will choose their actions for making a move according to a strategy (\definitionref{def:strategy}).
When the players execute their strategies in the game, we obtain a set of plays, called game outcomes. 
A strategy is winning if all the game outcomes are winning, no matter how the other player plays.

\begin{definition} \label{def:strategy}
 A \emph{strategy} for player $i$ in game $\game$ is a function $\sigma_i: \playpref(\game) \rightarrow \act_i$, such that $\sigma_i(\pi) \in \Gamma_i(\pi^q_{|\pi|-1})$ for any $\pi \in \playpref(\game)$.
 We write $\Sigma_i(\game)$ for the set of all player $i$ strategies in $\game$.
The \emph{outcome} of two strategies $\sigma_1 \in \Sigma_1(\game)$ and $\sigma_2 \in \Sigma_2(\game)$ is the set of plays that occur when Player 1 plays according to $\sigma_1$ and Player 2 according to $\sigma_2$:
\[\Outc(\sigma_1,\sigma_2) =  \{ \pi \in \Pi(\game) \mid \forall j \in \mathbb{N}: \   
\sigma_1(\pi_{0:j}) = \pi_{j+1}^a \wedge \sigma_2(\pi_{0:j}) = \pi_{j+1}^x \}\]

The plays occurring for strategy $\sigma_1 \in \Sigma_1(\game)$ are: 
\[\Outc(\sigma_1) \defis \bigcup\{\Outc(\sigma_1,\sigma_2) \mid \sigma_2 \in \Sigma_2(\game) \}\]
Strategy $\sigma_1$ is \emph{winning} with respect to reachability goal $R \subseteq Q$, if $\Outc(\sigma_1) \subseteq \winp(\game,R)$.
$\wins_1(\game,R)$ denotes all winning player 1 strategies.
Game $\game$ is \emph{winning} for player 1 w.r.t. goal $R$ iff $\wins_1(\game,R)\neq\emptyset$.
\end{definition}


\section{Model-based testing}
\label{sec:MBT}

Model-based testing is a smart way of testing that improves test efficiency by automatic test case generation, and execution, from a specification model, such that much manual (repetitive) labour can be prevented.
The specification is given as an automaton with input and outputs $\A$, that describes a restriction of desired system behavior.
The goal of model-based testing is then to determine whether a given System Under Test (SUT) behaves as described by its specification, i.e. where the SUT \emph{conforms} to $\A$.

\figureref{fig:MBT} illustrates the model-based testing process. 
A test case generation algorithm derives a set of test cases from the specification.
Here, test cases are finite scenarios composed from inputs and outputs. 
Then, these test cases are executed automatically  on the SUT.
The SUT is considered as a \emph{black box}, since we only see its input/output behavior, but not the internal workings, or source code.
Finally, every test case is assigned a \emph{test verdict}, being {\pass} or {\fail}. 
This verdict is determined in accordance with a conformance relation; 
in \sectionref{sec:alternating} we consider the input/output conformance relation \emph{ioco} \cite{tretmans,TBS11}. 




\newcommand{\model}[1]{
\begin{tikzpicture}[node distance=1.156cm,>=stealth']
	  \tikzstyle{every state}=[draw=black,text=black,inner sep=1pt,minimum
	  size=10pt,initial text=]
	  \node[state] (1) {};
	  \node[state,initial,initial where=right] (0) [below right=1] {};
	  \node[state] (3) [above right of=0] {};
	  \node [below=1mm of 0] {#1};
	  \path[->]
	  (0) edge [bend left] node {} (1)
	  (1) edge [bend left] node {} (0) 
	  (0) edge node {} (3)
	  (1) edge node {} (3)
	  (3) edge [loop below] node {} (3)
	  ;
      \end{tikzpicture}
}
\def\aboxr[#1,#2,#3,#4,#5]#6{%
  \node[draw, cylinder, alias=cyl, shape border rotate=90, 
  aspect=#3, %
  minimum height=#1, 
  minimum width=#2, 
  outer sep=-0.5\pgflinewidth,
  color=black] (#4) at #5 {};%
  \node at #5 {#6};%
  \fill [white!20] let \p1 = ($(cyl.before top)!0.5!(cyl.after top)$), \p2 =
  (cyl.top), \p3 = (cyl.before top), \n1={veclen(\x3-\x1,\y3-\y1)},
  \n2={veclen(\x2-\x1,\y2-\y1)} in (\p1) ellipse (\n1 and \n2); }

  \def\foldedpaper#1{
    \tikz[scale=#1,line width={#1*1pt}]{
        \def\a{1.41} 
        \def\b{0.2}  
        \def\c{0.1}  
        \def\d{0.05} 
        \def\N{6}    
        \draw         (0,0)
                --  ++(-1,0)
                --  ++(0,\a)
                --  ++(1-\b,0)
                --  ++(\b,-\b)
                -- cycle;
        \foreach \lnum in {1,...,\N}{
            \pgfmathsetmacro\yline{\a-\d-\lnum*\a/(\N+1)}
            \draw (-1+\c,\yline) -- (-\c,\yline);
        }
        \draw[fill=white] (0,\a-\b) -- ++(-\b,0) -- ++ (0,\b);
    }
}

\begin{figure}
\begin{subfigure}{0.45\textwidth}
\centering
  \begin{tikzpicture}
    \small
    \aboxr[40,50,1.4,tests,(0,0)] {Test cases};
    \node[fill=black, minimum height=2cm, minimum width=2cm,text=white, text width=1cm, align=center] (sut) at (0,-4){System Under Test};
    \node (model) at (4.5,0) {\model{Specification}};
    \node (verdict) at ($(tests)+(3.4,-1.8)$) {pass/fail};
    \path[shorten <= 2mm, shorten >= 2mm, ->, >=stealth]
        (tests) edge [line width=3mm] node [right] {Test execution} (sut)
        ($(tests)+(0,-0.62)$) edge [line width=2mm,white] ($(sut)+(0,1.11)$);
    \path[shorten <= 2mm, shorten >= 2mm, ->, >=stealth]
        ($(model)+(-1.11,0)$) edge [line width=3mm] node [below=2mm] {Test generation} (tests)
        ($(model)+(-1.16,0)$) edge [line width=2mm,white] ($(tests)+(0.97,0)$)
        ;
    \path[->]
    ($(tests)+(2.4,-1.8)$) edge node {} (verdict);
    \path[<->, shorten <=2mm, >= stealth,dashed]
	($(sut)+(1.2,-0.3)$) edge [line width=0.5mm, bend right=45] node [below right, text width=2cm, align=center] {Conformance relation} ($(model)+(0,-1)$);
  \end{tikzpicture}
  \caption{Model-based testing}
  \label{fig:MBT}
\end{subfigure}
 \begin{subfigure}{0.52\textwidth}
\centering
\begin{tikzpicture}[node distance=3cm,>=stealth']
\tikzstyle{every state}=[draw=black,text=black,inner sep=1pt,minimum size=13pt,initial text=]
\node[state] (1) [initial,initial where=above,initial text=] {$q_0$};
\node (2) [state, right of=1] {$q_1$};
\node (3) [state, below of=1] {$q_2$};
\node (4) [state, right of=3] {$q_3$};
\path[->]
(1) edge [bend left=60] node [above] {\play} (2)
(1) edge node [right] {\repeat} (3)
(1) edge [loop left] node [left] {$\delta$} (1)
(2) edge node [above] {\playList} (1)
(2) edge [loop right] node [right] {\song} (2)
(2) edge [bend left] node [below] {\quit} (1)
(2) edge node [left] {\repeat} (4)
(3) edge [bend left] node [left] {\quitRepeat} (1)
(3) edge node [above] {\play} (4)
(3) edge [loop left] node [left] {$\delta$} (3)
(4) edge [bend left] node [below] {\quit} (3)
(4) edge [bend right] node [right] {\quitRepeat} (2)
(4) edge [loop right] node [right] {\song} (4)
;
\end{tikzpicture}
\caption{SA specification of an MP3 player.
}
\label{fig:mp3player}
\end{subfigure}
\end{figure}


\subsection{Specifications as suspension automata}

Following~\cite{ncompletepaper}, we use suspension automata (SAs) as system specifications. These are determinised variants of the labeled transition systems with inputs and outputs from \cite{tretmans,STS13}.

For a partial function $f : X \rightharpoonup Y$, we write $f(x)\mydownarrow$ to denote that $f(x)$ is defined, and $f(x)\myuparrow$ to denote that $f(x)$ is undefined.

\begin{definition} \label{def:sa}
 A \emph{suspension automaton} (SA) is a 5-tuple $\A = (Q,L_I,L_O^\delta,T,q_0)$ where
 \begin{itemize}
  \item $Q$ is a non-empty finite set of states,
  \item $L_I$ is a finite set of input labels,
  \item $L_O^\delta \ = L_O \cup \{\delta\}$ with $L_O$ a finite set of output labels, $\delta \not\in L_O$, and $L_I \cap L_O^\delta = \emptyset$,
  \item $T : Q \times (L_I \cup L_O^\delta) \rightharpoonup Q$ is a partial transition function
  , and
  \item $q_0 \in Q$ is an initial state.
 \end{itemize}
We write $L \defis L_I \cup L_O^\delta$. 
For $q \in Q$, we denote the set of enabled inputs and outputs in $q$ by $\inp{q} = \{a \in L_I \mid T(q,a) \mydownarrow\}$, and $\out{q} = \{x \in L_O^\delta \mid T(q,x) \mydownarrow\}$ respectively.
We require that an SA is non-blocking: $\forall q \in Q: \out{q} \neq \emptyset$.
\end{definition}

We assume that any SA uses a special output label $\delta$ to indicate \emph{quiescence}, i.e., the absence of an observable output $x \in L_O$ \cite{tretmans}.
Handling quiescence is crucial in testing: if the SUT does not respond with any output, we must know whether or not this is allowed by the specification, otherwise we cannot come up with the correct verdict.
We formalized this with the non-blocking requirement in \definitionref{def:sa}.

\begin{example}
 \figureref{fig:mp3player} shows a specification of an MP3 player as an SA. 
In initial state $q_0$, no songs are played. Hence, this state has a self-loop labeled $\delta$. After a {\play} input, the system moves to state $q_1$, in which songs are being played, until either {\playList} occurs, or the {\quit} button is pressed.  
The MP3 player also features a repeat function, which can be switched on and off via the {\repeat} and {\quitRepeat} actions respectively. Thus, in state $q_3$, songs are played continuously, until the {\quit} action occurs. \end{example}

\subsection{Test assumptions in case of input-output conflicts}
SAs may feature states that enable both inputs and (non-quiescent) output actions. We call such a state \emph{mixed}: formally, $q\in Q$ is mixed if $\inp{q} \neq \emptyset$ and $\out{q} \neq \{\delta\}$. 
States $q_1$ and $q_3$ in \figureref{fig:mp3player} are mixed. 
Mixed states may give rise to \emph{input-output conflicts}. If  the tester wants to take an input action, and the SUT wants at the same time to take an output action, the  question arises which of the actions will be carried out.
The literature introduced different ways to handle input/outputs conflicts, i.e. \emph{test assumptions} on the interaction between the tester and the SUT.
Note that there is no `best' test assumption; this depends on the `hostility' of the SUT against the tester.
We list four test assumptions below.

\begin{itemize}
 \item A test interaction is \emph{input-eager (IE)} if the tester is always able to provide an input, even when the SUT wants  to produce an output.
 This assumption prioritizes inputs over outputs and thereby makes mixed states fully controllable for the tester. This underlies the framework in \cite{iotsioco}.
 \item The converse of input-eager is an \emph{output-eager (OE) } test interaction: the SUT always produces an output, unless $\delta$ is the only possible output.
 The authors of \cite{outputeager} use such an assumption.
This assumption prioritizes outputs over inputs and thereby makes mixed states fully uncontrollable for the tester.
 \item A test interaction is \emph{nondeterministic (ND)} if it is determined nondeterministically whether the SUT is able to take an output transition, or the tester to take an input transition in a mixed state.
 No guarantees are given on whether the tester is able to take an input transition in a mixed state, though this is not excluded as with the output-eager assumption.
 This assumption is similar to the test interaction used in the original theory for labeled transition systems with ioco \cite{tretmans}.
 \item A test interaction is \emph{input fair (IF)} if the tester is eventually able (after trying finitely many times) to take any input transition in a mixed state.
 Hence, mixed states are controllable for the tester, at the expense of trying multiple times.
 This assumption is made in \cite{ncompletepaper}.
\end{itemize}

In all test interactions of the assumptions above, either an input or an output action is ignored. 
One could also take both into account, by executing them both, but in nondeterministic order. 
This is a concurrent interpretation of an input-output conflict, which may be well suited for systems dealing with concurrent processes.
If the SUT is able to receive more inputs than its specification specifies, this may lead to different interpretations of this assumption.
Therefore, we will delay formalizing this assumption to future work.
All test assumptions from the list above are formalized in \sectionref{sec:iots2game}, by incorporating the assumption in the $\moves$ function from the underlying game arena of the specification.

\begin{example}
An MP3 player, as described in \figureref{fig:mp3player}, does not function properly, when taking the output-eager or nondeterministic test assumption. A real world MP3 player normally responds to input (though there may be some delay).
With the printer from \figureref{fig:printer}, we show that all test assumptions can impose a useful interpretation on the test interaction between the tester and the SUT.
This specification models a printer which can handle printing and scanning in an interleaved way, i.e. printing does not need to be finished before scanning to be started and vice versa.
 \begin{itemize}
 \item The input-eager assumption allows the tester to always provide the inputs {\print} and {\scan}, before receiving outputs {\printed} and {\scanned}. Only if the tester decides to wait for an output, the SUT can produce these.
 
 \item  The output-eager test assumption expresses that the specification is too complicated for the SUT that is being tested: the SUT cannot print and scan at the same time, because a {\printed} or {\scanned} output will occur after the tester has sent the respective input.
 
 \item  With the nondeterministic test assumption, the tester may succeed in sending both inputs {\print} and {\scan} before receiving outputs, but no guarantees on success can be given.
 Furthermore, providing a {\print} input in mixed state $q_1$ may result in taking this transition, but instead, the {\scanned} output transition may also be taken.
 \item The input-fair test assumption is similar to the nondeterministic test assumption, but the difference is that it guarantees that the two input transitions {\print} and {\scan} can be taken from $q_1$ and $q_3$, before the outputs are produced, after trying a few times.
  \end{itemize}
\end{example}

\vspace{-8mm}
 \begin{figure}[ht!]
   \begin{subfigure}{0.4\textwidth}
 \hspace{-1.1cm}
\begin{tikzpicture}[node distance=2.2cm,>=stealth']
 \tikzstyle{every state}=[draw=black,text=black,inner sep=1pt,minimum
        size=10pt,initial text=]
  \node[state,initial,initial where = above] (1) {$q_0$};
  \node[state] (2) [fill=black,below of=1, text=white] {$q_3$};
  \node[state] (3) [below of=2] {$q_6$};
  \node[state] (4) [fill=black,right of=1, text=white] {$q_1$};
  \node[state] (5) [right of=4] {$q_2$};
  \node[state] (6) [right of=2] {$q_4$};
  \node[state] (7) [right of=6] {$q_5$};
  \node[state] (8) [right of=3] {$q_7$};
  \path[->]
  (1) edge node [right] {\print} (2)
  (2) edge node [right] {\printed} (3)
  (4) edge node [right] {\print} (6)
  (6) edge node [right] {\printed} (8)
  (5) edge node [left] {\print} (7)
  (7) edge [in=35,out=45,looseness=2] node [above right=0.9mm and -4mm] {\printed} (1)
  (1) edge node [above] {\scan} (4)
  (4) edge node [above] {\scanned} (5)
  (2) edge node [below] {\scan} (6)
  (6) edge node [below] {\scanned} (7)
  (3) edge node [below] {\scan} (8)
  (8) edge [in=235,out=225,looseness=2] node [below left= 0.9cm and -1.3cm] {\scanned} (1)
  (1) edge [loop left] node [left] {$\delta$} (1)
  (3) edge [loop left] node [above left=1mm and -3mm] {$\delta$} (3)
  (5) edge [loop right] node [below right=1mm and -3mm] {$\delta$} (5)
  ;
 \end{tikzpicture}
 \vspace{-1.5cm}
 \caption{SA specification of a printer.
 Mixed states are black.}
 \label{fig:printer}
 \end{subfigure}
 \quad
 \begin{subfigure}{0.57\textwidth}
\centering
\begin{tikzpicture}[node distance=2.5cm,>=stealth']
\tikzstyle{every state}=[draw=black,text=black,inner sep=1pt,minimum size=10pt,initial text=]
\node[state] (1) [initial,initial where=left,initial text=] {$t_0$};
\node (2) [below right of=1] {\fail};
\node (3) [below left of=1] {\fail};
\node (4) [state, below of=1] {$t_1$};
\node (5) [state,below right of=4] {$t_3$};
\node (6) [state, below left of=4] {$t_2$};
\node (7) [below of=4] {\fail};
\node (8) [below of=5] {\fail};
\node (9) [below left of=5] {\pass};
\node (0) [below right of=5] {\fail};
\node (10) [below left of=6] {\pass};
\node (11) [below of=6] {\pass};
\node (12) [left of=4] {\fail};
\node (13) [left of=6] {\fail};
\path[->,>=stealth]
(1) edge[dashed] node [above right] {\printed} (2)
(1) edge[dashed] node [above left] {\scanned} (3)
(1) edge node [below right] {\print} (4)
(4) edge[dashed] node [above right] {\printed} (5)
(4) edge node [left] {\scan} (6)
(4) edge[dashed] node [below left=2mm and -1mm] {\scanned} (7)
(5) edge node [below right=1mm and -1mm] {\printed} (8)
(5) edge [right] node {$\delta$} (9)
(5) edge node [right=1mm] {\scanned} (0)
(6) edge node [above left] {\scanned} (10)
(6) edge node [right] {\printed} (11)
(4) edge[dotted] node [above] {$\delta$} (12)
(6) edge node [above] {$\delta$} (13)
;
\end{tikzpicture}
\caption{Test case $\T$ of the SA of \figureref{fig:printer}.
For readability, {\pass} and {\fail} are displayed multiple times, and their self-loops have been omitted. See \exampleref{exmp:testcase} for dotted and dashed transitions.}
\label{fig:testcase}
 \end{subfigure}
\end{figure}
\vspace{-5mm}

\subsection{Test cases} 

We give a definition of test cases in the spirit of \cite{tretmans}.
As shown in \figureref{fig:testcase}, a test case is a finite and acyclic SA $\T$.
A test case is constructed by repeatedly taking either of the following test steps:
\begin{enumerate}
\item Choose an input $a?$ from the input actions enabled in the current specification state, execute the input action, and move to the next state in $\T$. If the specification is in a mixed state, then it may happen that an output was observed before $a?$ was observed. Therefore, if $a?$ is enabled in a state of the test case, then all outputs from $L_O$ are enabled as well. 
\item Observe an output from the SUT.
In case an output is observed that is prohibited by the specification, a fail verdict is emitted. Otherwise, one moves to the next state in $\T$. 
\item Stop testing and emit a pass verdict. Note that all fail verdicts are handled in step 2.
\end{enumerate}

Before defining test cases formally, we first define some auxiliary notation. 

 \begin{definition} \label{def:iocooperations}
 Let $\A = (Q,L_I,L_O^\delta,T,q_0)$ be an SA, $q \in Q$, $Q' \subseteq Q$, $\mu \in L$, $\rho \in L^*$, and $\epsilon$ the empty sequence. Then we define:

 \vspace{-1em}
 \noindent
 \begin{minipage}[t]{0.5\textwidth}
\small
 \begin{align*}
 &\afterm{q}{\epsilon} = \{q\}\\
 &\afterm{q}{\mu\rho} = 
   \begin{cases}
     \after{T(q,\mu)}{\rho} & \text{if } T(q,\mu) \mydownarrow\\
     \emptyset & \text{otherwise}
   \end{cases}\\
    &\afterm{\A}{\rho} = \afterm{q_0}{\rho}\\
 \end{align*}
\end{minipage}
\begin{minipage}[t]{0.4\textwidth}
\small
 \begin{align*}
 & \outm{Q'} = \bigcup_{q' \in Q'}\outm{q'}\\
 & \stracesm{\A} = \{\rho' \in L^* \mid \afterm{\A}{\rho'} \neq \emptyset\}\
 \end{align*}
\end{minipage}
\end{definition}

\begin{definition} \label{def:testcase}
 A \emph{test case} for an SA $\A$ is an SA $\T=(Q^t,L_I,L_O^\delta,T^t,q_0^t)$ such that:

 \begin{itemize}
  \item There are two special states \pass, \fail $\in Q^t$ such that $T^t(\pass,x) = \pass$ and $T^t(\fail,x) = \fail$ for all $x \in L_O^\delta$, and  $T^t(\pass,a) = T^t(\fail,a)\myuparrow\ $ for $a \in L_I$.
  \item $\T$ has no cycles except those in $\pass$ and $\fail$.
  \item Every state enables all outputs $L_O$, and either one input 
  (matching step 1 above)   or $\delta$ (matching step 2):
  $\forall q \in Q: (|\inp{q}| = 0 \wedge \out{q} = L_O^\delta) \vee (\out{q}=L_O \wedge |\inp{q}|=1)$.
  Exception: in case of an input-eager test interaction, we only require: $\forall q \in Q: (|\inp{q}| = 0 \wedge \out{q} = L_O^\delta) \vee |\inp{q}| = 1$.
  \item Traces of $\T$ leading to {\pass}, are traces of $\A$, while traces to {\fail} are not.\\
  $\forall \rho \in \straces{\T}: (\after{\T}{\rho} = \pass \implies \rho \in \straces{\A}) \wedge (\after{\T}{\rho} = \fail \implies \rho \not\in \straces{\A})$
 \end{itemize}
\end{definition}

\begin{example} \label{exmp:testcase}
\figureref{fig:testcase} shows a test case for the printer specification of \figureref{fig:printer}.
In state $t_0$, the tester provides the input {\print}. 
This state also has all non-quiescence outputs {\printed} and {\scanned}, because some test assumptions allow these to occur instead of the SUT accepting input {\print}. 
Since none of these actions are allowed by the specification in \figureref{fig:printer}, these actions lead to a $\fail$ verdict. 
In state $t_1$, the tester provides input {\scan}, but now one of the non-quiescent outputs is allowed, namely {\printed}.
In state $t_3$, the tester decides to observe an output from the SUT. She may see either of three things: (1) quiescence (i.e. $\delta$), leading to a $\pass$ verdict, since quiescence is allowed in the specification, (2) {\printed}, which is not allowed, and (3) {\scanned}, which is also not allowed.
After observing quiescence from state $t_3$, the tester decides to stop testing and conclude verdict \pass.

Test cases are constructed without taking into account any specific test assumption, by including all inputs and outputs relevant for any of them. The dashed output transitions of $t_0$ and $t_1$ (and all of $t_3$) can be omitted in case of the input-eager test assumption. 
In case of the output-eager test assumption, input \{\scan\} cannot be taken after \{\print\}, because specification state $T(q_0,\print)= q_3$ is mixed.
One can adapt test case $\T$ to be relevant for the output-eager test assumption, by
omitting the {\scan} transition from $t_1$ (and all of $t_2$), while adding the dotted $\delta$ transition in $t_1$ for satisfying the third rule of \definitionref{def:testcase}.
\end{example}

\section{Specifications are game arenas}
\label{sec:iots2game}

To study the connection between test cases and games, we associate to each specification SA $\A$ a game arena $\game_\A$. 
In $\game_\A$, the tester (player 1) and the SUT (player 2) play on the state space given by $\A$, extended with a sink state $\stopState$, and a number $i \in \{1,2\}$ indicating whether the state was reached via a player 1 (input) or player 2 (output) action. The latter is important, because testers see the SUT via their traces, so we must record whose action was carried out. 

To advance the game $\game_\A$, both the tester and the SUT choose an action from the current state:
\begin{itemize}
\item 
The tester chooses either an enabled input from the specification SA $\A$, or one of the special inputs $\theta$ and $\stop$.
The $\theta$ action expresses that the tester desires to take no input, and allows the SUT to execute any output he wishes; the $\stop$ action indicates that the tester wants to stop testing, which brings the game to the state $\stopState$.
\item 
The SUT chooses one of the enabled outputs from the specification.
\end{itemize}
Then the game moves to a next state, according the function $\moves$, which reflects how the tester and the SUT interact.
Hence, different test assumptions made in the literature give rise to different definitions of the $\moves$ function. 

The explicit game definition (\definitionref{def:ioltstogame}) and encoding of test assumptions (\sectionref{sec:testassumptions}) set our work apart from earlier game-based approaches to testing.
In particular, \cite{coveragecomplexity} derive optimal test cases from a specification that is already given as a game. 
The encoding also allows us to study the relation between test cases and games strategies (\sectionref{sec:teststrategy}), and between alternating refinement and conformance (\sectionref{sec:alternating}).

\begin{definition} \label{def:ioltstogame}
Let $\A=(Q,L_I,L_O^\delta,T,q_0)$ be an SA. The  \emph{game arena underlying $\A$} is defined by 
$\game_\A = (\qgame,(q_0,1),\act_1,\act_2,\Gamma_1,\Gamma_2,\moves)$ where:
 \begin{itemize}
 \item $\qgame = (Q\times \{1,2\}) \cup \{(\stopState,1)\}$
  \item $\act_1 = L_I\cup \{\theta, \stop\}$ and $\act_2 = L_O^\delta$,
  \item for all $q \in Q$ and $i \in \{1,2\}$, we take $\Gamma_1((q,i)) =  \inp q\cup\{\theta,\stop\} $ and $\Gamma_2((q,i)) = \out{q}$, 
  \item we take $\Gamma_1(\stopState, 1) = \{\stop\}$ and $\Gamma_2(\stopState, 1) = L_O^\delta$, and 
  \item the function $\moves: \qgame \times Act_1 \times Act_2 \rightarrow 2^{\qgame}$ encodes one of the different test assumptions and is given in Subsection~\ref{sec:testassumptions}.
  Besides the requirement from \definitionref{def:gamearena} for moves with undefined action, we require $\moves(q,\stop,x) = (\stopState,1)$, and $\moves((\stopState,1), \stop, x) = (\stopState,1)$.
  \end{itemize}
\end{definition}

For the remainder of the paper, we fix a specification $\A$, and its underlying game $\game_\A$.

\subsection{Encoding test assumptions}
\label{sec:testassumptions}

We formalize test interaction 
by implementing a {\moves} function for each of the test assumptions.
Note that the {\moves} function is of type $ \qgame \times (L_I \cup \{\theta, \stop\}) \times L_O^\delta \rightarrow 2^{\qgame}$, i.e. it takes a game state, an input from the tester, and an output from the SUT. The function returns a set of the reached next states for the given input and output, using the transition function of the SA for which this \emph{moves} function is defined.


All the {\moves} functions from \definitionref{def:moves} use the symbols $\delta$ and $\theta$ as special symbols that transfer control to the tester or SUT respectively.
The symbol $\delta$ is used in its usual semantics, i.e. it denotes quiescence. When the SUT is quiescent, the tester is always able to provide an input. 
However, $\delta$ can only actually be observed if the tester chooses the $\theta$ input.
Hence, $\theta$ is then used as  an artificial input to model that the tester is waiting for the SUT to take an output transition.
This corresponds with how $\theta$ is used in \cite{tretmans}.

In practice, $\delta$ can be observed by setting a timeout.
It is then assumed that the SUT does not produce any regular output from $L_O$ after this timeout.
Input $\theta$ can be implemented in practice, by waiting for the SUT to produce an output.

The behaviour of the {\moves} functions from \definitionref{def:moves} differ for the regular inputs $\inp{q}$ and outputs $\out{q}\setminus\{\delta\}$ of a state $q$ of the SA, in order to resolve input/output conflicts according to one of the test assumptions from \sectionref{sec:MBT}.
We will discuss how to implement the input-fair test assumption in \sectionref{sec:fairness}, because its semantics cannot be implemented directly with only a {\moves} function.

\begin{definition} \label{def:moves}
Let $\A=(Q,L_I,L_O,T,q_0)$ be an SA. The various test assumptions from \sectionref{sec:MBT} give rise to the following functions $\moves: \qgame \times (L_I \cup \{\theta, \stop\}) \times L_O^\delta \rightarrow 2^{\qgame}$ (we do not include the moves from \definitionref{def:ioltstogame} for $(\stopState,q)$, $\stop$ and undefined actions here):
\begin{itemize}
\item 
In the input-eager regime, an input $a\in L_I$ is always executed, unless the tester decides to not perform an input, i.e. she proposes $a=\theta$.
\begin{align*}
\moves_{IE}((q,i),a,x) = 
\begin{cases}
  \{(T(q,a), 1)\} &\mbox{if } a \neq \theta \\
  \{(T(q,x), 2)\} & \mbox{otherwise } 
\end{cases}
\end{align*}
\item 
In the output-eager regime, the output  action $x \in L_O$ is always executed, 
unless both $x=\delta$ and $a\neq\theta$.
\begin{align*}
\moves_{OE}((q,i),a,x) =
\begin{cases}
  \{(T(q,x), 2)\} & \mbox{if } x \neq \delta\ \vee\ a = \theta\\
  \{(T(q,a), 1)\} &\mbox{otherwise}
\end{cases}
\end{align*}
\item In the nondeterministic regime, a nondeterministic choice is made whether to execute the input (unless $a=\theta$) or the output (unless $x=\delta$ and $a\neq\theta$). We take the same {\moves} function for the input-fair regime, and explain this in \sectionref{sec:fairness}.
\begin{align*}
\moves_{ND}((q,i),a,x) = 
\begin{cases}
  \{(T(q,x),2)\} &\mbox{if } a = \theta \\
  \{(T(q,a),1)\} & \mbox{if } x = \delta \\ 
  \{(T(q,a),1), (T(q,x),2)\} &\mbox{otherwise }
\end{cases}
\end{align*}
\end{itemize}
\end{definition}

\paragraph{\bf Testing with a hard reset.}
Further, it is often useful to have a hard reset function. 
In testing, a reset function is used to execute multiple test cases from the initial state. 
In the specification, a hard reset is enabled in any state, and the tester can always use it to go back to the initial state, no matter whether the SUT wanted to do an output action. 
In practice, a SUT often needs to be instrumented for implementing such a reset function in a fast way, as rebooting a system can take a lot of time.
In \definitionref{def:ioltsgamewithreset}, we have adapted \definitionref{def:ioltstogame}, to include this special input $\reset \not\in L_I$.

\begin{definition} \label{def:ioltsgamewithreset}
Let $\A=(Q,L_I,L_O^\delta,T,q_0)$ be an SA.
The  \emph{resettable game arena underlying} $\A$ is defined by 
$\game_\A^\reset = (\qgame,(q_0,1),\act_1,\act_2,\Gamma_1,\Gamma_2, \moves)$ where:
 \begin{itemize}
  \item $\qgame = (Q\times \{1,2\}) \cup \{(\stopState,1)\}$
  \item $\act_1 = L_I\cup \{\theta,\stop,\reset\}$ and $\act_2 = L_O^\delta$
  \item for all $q \in Q$ and $i \in \{1,2\}$, we take  $\Gamma_1((q,i)) =  \inp q\cup\{\theta,\stop,\reset\}$ and $\Gamma_2((q,i)) = \out{q}$,
  \item we take $\Gamma_1(\stopState, 1) = \{\stop\}$ and $\Gamma_2(\stopState, 1) = L_O^\delta$, and 
  \item the function $\moves: \qgame \times Act_1 \times Act_2 \rightarrow 2^{\qgame}$ encodes one of the test assumptions from \definitionref{def:moves}, with the modification that $\forall q \in Q, \forall i \in \{1,2\}, \forall x \in \Gamma_2((q,i)): \moves((q,i),\reset,x)=(q_0,1)$.
  \end{itemize}
\end{definition}

\subsection{Encoding the input-fair test assumptions via fair plays}
\label{sec:fairness}

To define the input-fair test assumption from \sectionref{sec:MBT}, we introduce the notion of a \emph{fair play}.
In essence, a play $\pi$ is input fair if, whenever the tester wants to perform an input action from state $q$, it is eventually executed in $q$ \cite{ncompletepaper}.
Given a state $q$ and an input action $a$ that is enabled in $q$, we say that a play is \emph{fair with respect to $q$ and $a$} if $a$ is actually taken in $q$ at some point in time. 
A play $\pi$ is \emph{fair} if it is {fair} with respect to any $q$ appearing in $\pi$ and any input action that was proposed in $q$ in $\pi$.

\begin{definition} \label{def:fairness}
 Let $\pi \in \Pi(\game_\A)$. Then $\pi$ is \emph{input-fair} w.r.t. some state $q \in Q$ and some input $a \in \inp{q}$ if:
  \begin{align*}
   \exists j \in \mathbb{N}, i \in \{1,2\}, q' \in Q: \pi_j^a = a \wedge \pi_j^q = (q,i) \wedge \pi_{j+1}^q = (q',1)
  \end{align*}
 Play $\pi$ is \emph{input-fair} if:
 \begin{align*}
  \forall q \in Q, a \in L_I: (\exists j \in \mathbb{N}, i \in \{1,2\}: \pi_j^a = a \wedge \pi_j^q = (q,i)) \implies \pi \textnormal{ is input-fair w.r.t. } q \textnormal{ and } a
 \end{align*}
\end{definition}

Clearly, restricting an underlying game arena of an SA with the nondeterministic test assumption to input-fair plays only results in  satisfying the description of the input-fair test assumption from \sectionref{sec:MBT}.
Instead of restricting the plays of the game, we will leave the game with the nondeterministic test assumption as is, and consider input-fair strategies in \definitionref{def:inpfstrat}.

\begin{definition} \label{def:inpfstrat}
 Let $\game_\A^{ND}$ be the underlying game arena of SA $\A$ with the nondeterministic $\moves$ function from \definitionref{def:moves}. Then a strategy $\sigma \in \Sigma_i(\game_\A)$ is \emph{input-fair} if it is winning on the set of all input-fair plays of $\game_A$.
\end{definition}

Requiring input-fairness may appear restrictive for specifications with states that cannot be reached in any way once leaving them via some transition.
However, adding a reset to the game (\definitionref{def:ioltsgamewithreset}) resolves this, as a state reached by some play, can then always be reached again by resetting, and following the same play.
Of course, the SUT may also be able to prevent the tester from following this play right away, but input-fairness ensures that this is possible after trying a few times.

\subsection{Comparing strategies for different test assumptions}

This section compares strategies for game arenas based on the different test assumptions, constructed from the same specification. 
In particular, we show that, for any reachability objective $R$,
either both the input-eager and input-fair games are winning for player 1, or both not.
Similarly, for objective $R$, either both the nondeterministic and output-eager games are winning for player 1 or both not.

To show that the input-eager and input-fair games coincide, we observe that any winning player 1 strategy of an input-fair game is also winning in an input-eager game.
The converse does not hold however, because the set of plays of an input-eager game has fewer plays than the input-fair game, as the input-fair game does not have plays in which a proposed input transition is not taken.
However, by reaching the same state multiple times, we know that the winning action proposed in the input-eager game is eventually carried out.

A similar reasoning holds for the winning player 1 strategies of games based on the nondeterministic and output-eager test assumption.
There are no guarantees for taking an input transition in a mixed state (according to the nondeterministic test assumption), or it is simply not possible (according to the output-eager test assumption).
Hence, in both cases, a winning strategy cannot make use of taking input transitions in mixed states.
Hence, the existence of winning strategies is the same in both games.

\begin{theorem} \label{thm:winsubset}
 Let $\A=(Q,L_I,L_O,T,q_0)$ be an SA, and $\game_\A^{IE}, \game_\A^{IF}, \game_\A^{OE}, \game_\A^{ND}$ the underlying game arenas of $\A$ for the input-eager, input-fair, output-eager, and nondeterministic test assumption, respectively. Let $R\subseteq Q$ be a reachability goal. 
 Then:
 \begin{enumerate}
 \item $\game_\A^{IE}$ is winning for player 1 w.r.t. $R$ if and only if $\game_\A^{IF}$ is winning for player 1  w.r.t. $R$, and
 \item $\game_\A^{OE}$ is winning for player 1 w.r.t $R$ if and only if $\game_\A^{ND}$ is winning for player 1 w.r.t. $R$.
\end{enumerate}
\end{theorem}

%

\section{Test cases are game strategies (and vice versa)}
\label{sec:teststrategy}

This section establishes a strong correspondence between Player 1 strategies and test cases.
To achieve this result, we observe that test cases and strategies share many features, but differ on two aspects: 
(1) Test cases are finite, while strategies play forever. Hence, test cases correspond to Player 1 strategies that are finite, i.e., eventually provide a $\stop$ action. 
(2) Test cases base their decisions on the observed traces only, while strategies can use all information contained in the plays, especially the proposed actions for which the corresponding transition was not taken.
Therefore, test cases correspond to finite trace-based Player 1 strategies.
Thus, we establish a bijection between test cases and finite, trace-based strategies. 
We first explain how a strategy can be extracted from a test case, and then how a test case can be extracted from a strategy.


\subsection{From strategies to test cases}

We derive a test case from each finite and trace-based Player 1 strategy (\definitionref{def:tracebasedfinite}). 
A strategy is \emph{trace-based} if its choices only depend on the observed traces.
Finite strategies eventually provide a $\stop$ action. Note that, after one $\stop$ action from Player 1, all subsequent Player 1 actions are $\stop$. Hence, only the prefix before the $\stop$-action matters (\definitionref{def:strattotest}).

Exactly the traces of these prefixes cut off before the {\stop} actions can be used to construct a test case.
Each of these traces either leads to the {\pass} state, or to a unique test case state.
A test case only consisting of these traces does not contain any output transition to the {\fail} state, because these outputs do not occur in the specification. Hence, they need to be added.
This construction proves \theoremref{thm:strattotest}.

\begin{definition} \label{def:tracebasedfinite}
The function $\trace: \playpref(\game_\A)\to L^*$ assigns to each play prefix 
$\pi$
$ = (q_0,i_0) \langle a_0,x_0\rangle (q_1,i_1) \allowbreak \dots\langle a_{n-1}, x_{n-1}\rangle  (q_n,i_n)$, 
a sequence of action labels  $\rho = \rho_0\rho_1,\ldots \rho_{n-1}$ given by
$\rho_j = a_j$ if $i_{j+1}= 1$ and $\rho_j = x_j$ otherwise.
A strategy $\sigma\in\Sigma_i(\game_\A) $ is called \emph{trace-based} if: 
\[\forall \pi,\pi'\in\playpref(\game): {\trace(\pi)=\trace(\pi')} \implies \sigma(\pi)=\sigma(\pi')\]

A strategy $\sigma_1 \in \Sigma_1(\game_\A)$ is \emph{finite} if: \[\forall \pi \in 
\Outc(\sigma_1), \exists j\in \mathbb{N}: \pi^a_j =\stop\]
\end{definition}

\begin{example}
 A strategy is not trace-based, if it returns different actions for a play that consists of the same executed actions that form the trace, and different non-executed actions.
 This situation cannot occur for the printer from \figureref{fig:printer}, so we give an example player 1 strategy for the MP3 player of \figureref{fig:mp3player}:
 \vspace{-3mm}
 \[
  \sigma_1(\pi) =
  \begin{cases}
   \play &\mbox{if } \pi = (q_0,1) \langle \play, \delta \rangle (q_1,1) \langle \quit, \playList \rangle (q_0,1)\\
   \repeat &\mbox{if } \pi = (q_0,1) \langle \play, \delta \rangle (q_1,1) \langle \quit, \song \rangle (q_0,1)\\
   \theta &\mbox{otherwise}
  \end{cases}
\]
 State $q_1$ enables two outputs (namely {\song} and {\playList}) which are both not executed, in the two plays mentioned above, because the input transition for {\quit} has been taken (as indicated by the 1 in the last state).
 Nevertheless, $\sigma$ returns either {\play} or {\repeat} based on the non-executed outputs.
\end{example}

\begin{definition} \label{def:strattotest}
  Let $\sigma_1 \in \Sigma_1(\game_\A)$ be a player 1 trace-based, finite strategy in $\game_\A$. 
  We define a trace set $T_{\sigma_1} = \{\trace(\pi) \mid \pi \in \pref(\Outc(\sigma_1)) \wedge  \pi_{|\pi|-1}^a \neq \stop\}$. 
  \end{definition}

\begin{theorem} \label{thm:strattotest}
Then $T_{\sigma_1}$ characterizes a unique test case. 
\end{theorem}

\begin{example} \label{exmp:strattotest}
 Let $\sigma_1 \in \Sigma_1(\game_\A)$ be defined as follows:\\
 $\sigma_1(\pi) = 
 \begin{cases}
  \stop &\mbox{if } |\pi|\ge 4\\
  \print &\mbox{if } |\pi| < 4 \wedge \print \in \Gamma_1(\pi^q_{|\pi|-1})\\
   \scan &\mbox{if } |\pi| < 4 \wedge \print \not\in \Gamma_1(\pi^q_{|\pi|-1}) \wedge \scan \in \Gamma_1(\pi^q_{|\pi|-1})\\
  \theta &\mbox{otherwise}
 \end{cases}$\\
 Note that strategy $\sigma_1$ is finite and trace-based.
 The trace set of $\sigma_1$ is $T_{\sigma_1} = \{\print, \print\printed, \allowbreak \print\printed\delta,\ \print\scan,\ \print\scan\printed,\ \print\scan\scanned\}$, in case of the input-fair or nondeterministic test assumption.
 This set is exactly the prefix-closed set of the traces leading to a {\pass} state in the test case of \figureref{fig:testcase}.
 Note that if $\game_\A$ uses the input-eager test assumption, traces $\print\printed$ and $\print\printed\delta$ are not included in $\T_{\sigma_1}$.
 The traces $\print\scan,\ \print\scan\printed$, and $\print\scan\scanned$ are not included in $\T_{\sigma_1}$ in case of the output-eager test assumption.
\end{example}

 \subsection{From test cases to strategies}

Given a test case $\T$ for an SA $\A$, we construct a game strategy $\sigma_1$ as follows. On play prefixes $\pi$ whose traces are included in $\T$, $\sigma_1$ returns the input action enabled in the state of $\T$ reached by this trace, if it has one. If $\pi$ has a trace in $\T$ leading to (or passing by) the {\pass} state, then $\sigma_1$ returns the action {\stop}. In all other cases, we set $\sigma_1(\pi)= \theta$, because the trace of $\pi$ then either reaches a state of $\T$ with no enabled input transition, or $\pi$ is a play that does not occur in the outcome of the game when using $\sigma_1$.

\begin{definition} \label{def:testtostrat}
Let $\T$ be a test case for an SA  $\A$. We define a strategy $\sigma_\T$ of $\game_\A$ as follows:
\begin{align*}
\sigma_\T(\pi) = 
\begin{cases}
  a     &   \mbox{if } \exists q \in Q, a \in L_I: \after{\T}{(\trace(\pi))} = \{q\} \wedge \inp{q} = \{a\} \\
  \stop & \mbox{if }  \exists j \in \mathbb{N}: \pass \in \after \T{(\trace(\pi_{0:j}))} \\
\theta &\mbox{otherwise}
\end{cases}
\end{align*}
\end{definition}

Note that $\sigma_\T$ is well-defined, because by \definitionref{def:testcase}, the input $a$ in the first clause is unique, if it exists. 
\theoremref{thm:testtostrat} then states that  $\sigma_\T$ is finite and trace-based Player 1 strategy. Further, $\sigma_T$ is unique, i.e. from a test case exactly one strategy can be derived.

\begin{theorem} \label{thm:testtostrat}
Let $\T$ be a test case for SA $\A$. Then we have
\begin{enumerate}
\item The strategy  $\sigma_\T$ is a Player 1 strategy in $\game_\A$.
\item $\sigma_\T$ is finite and trace-based.
\item If $\sigma_\T = \sigma_{\T'}$ then $\T=\T'$.
\end{enumerate}
\end{theorem}

\begin{example}
 We use \definitionref{def:strattotest} to construct the following strategy from test case $\T$ of \figureref{fig:testcase}:
 \begin{align*}
  \sigma_\T(\pi) = 
\begin{cases}
  \print     &   \mbox{if } \trace(\pi) = \epsilon \\
  \scan &\mbox{if } \trace(\pi) = \print\\
  \stop & \mbox{if } \trace(\pi) \in \{\print\scan\printed\stop^*,\print\scan\scanned\stop^*,\\
  & \qquad\qquad\qquad\quad\qquad\qquad\qquad\qquad\qquad\qquad\qquad \print\printed\delta\stop^*\} \\
\theta &\mbox{otherwise}
\end{cases}
 \end{align*}
 Note that this strategy is equivalent to the one from \exampleref{exmp:strattotest}.
\end{example}

\subsection{Test case generation is strategy synthesis}

We can now establish that we have defined a bijective function between test cases and strategies, by using the translation from strategy to test case from \theoremref{thm:strattotest}, and the translation from test case to strategy from \theoremref{thm:testtostrat} as its inverse. 

\begin{theorem} \label{thm:testcasestratbijection}
The function $\T \mapsto \sigma_\T$ is a bijection from the set of test cases of $\A$ to the set of finite trace-based strategies of $\game_\A$.
\end{theorem}

A consequence of \theoremref{thm:testcasestratbijection} is that  game synthesis algorithms can be used for deriving test cases for specific \emph{test objectives}.
Test objectives describe the objective that a tester likes to achieve during testing. 

Various test objectives exist:  \emph{Reachability goals} \cite{BrandanBB04} are states in the specification that the tester likes to reach. For example, one may like to see that the MP3 player is able to play songs. To do so, the tester likes to reach any state with an outgoing song-transition (and see if this transition can be executed), conforming to states $q_1$ and $q_3$ in \figureref{fig:mp3player}. Test purposes \cite{confrelandtestobj} generalize reachability goals in the sense that a whole scenario needs to be executed; for example, one likes to see if the MP3 player can produce songs, after a quit? action. Since such scenarios can be adaptive, 
 we model a test purpose as an SA with final states, in which the test purpose was successfully executed. This idea is common in model-based testing \cite{confrelandtestobj}, but has not been exploited in a game theoretic setting. 
 The interaction between the specification and the test purpose is modeled via a composition operator $\|$. Finally, (state) \emph{coverage} \cite{coveragecomplexity} can be a test objective, where the tester tries to cover as many states in the specification as possible. 
 As stated, our framework enables strategy synthesis for these test objectives for any test regime. 


\section{Conformance is alternating trace inclusion}
\label{sec:alternating}

A popular conformance relation for model-based testing is \emph{input-output conformance}, $\ioco$ for short \cite{tretmans}. 
This relation formalizes what it means that an SUT, modeled as an input-enabled suspension automaton $\impl$, conforms to a specification, modeled as an SA $\spec$ (\definitionref{def:ioco}); 
an SA $\A$ is input-enabled if all its states $q$ satisfy $ \inp{q} = L_I$.
The $\ioco$ relation allows the implementation to implement more inputs, and fewer outputs than the specification. 
Indeed, implementation $\impl$ may implement more services than specified in $\spec$, but on the specified inputs it must behave as prescribed by $\spec$.

This viewpoint corresponds to Player 2 alternating trace inclusion $\alttraceincl$ for games \cite{alternatingincl}. 
Game $\game_A$  is 2-alternating-trace included in game $\game_B$, if any trace set that can be enforced by Player 2 in $\game_A$ can also be enforced by Player 2 in $\game_B$ (\theoremref{thm:alttrincl}).

In the definition of alternating trace inclusion (\definitionref{def:alttrincl}), player 1 chooses an input in $\game_\B$, and then needs to choose a corresponding input in $\game_\A$.
However, we need to take care that player 1 does not cheat in $\game_\A$, by choosing the $\theta$ input, if this input is not chosen in $\game_\B$ (\definitionref{def:cheat}).

\begin{definition}
 \label{def:ioco}
 Let $\A$ and $\B$ be SAs over the same label sets and assume that $\A$ is input-enabled. 
 Then we say that $\A \ioco \B$ if for all $\rho \in \straces{\B}$ we have $\out{\after{\A}{\rho}} \subseteq \out{\after{\B}{\rho}}$.
\end{definition}

\begin{definition} \label{def:cheat}
 Let $\game$ be a game arena corresponding to some SA, and $\pi = (q_0,j_0)\langle a_0,x_0 \rangle (q_1,j_1) \langle a_1, x_1 \rangle \allowbreak (q_2,j_2) \dots \langle a_{k-1} x_{k-1} \rangle (q_k,j_k) \in \playpref(\game)$ a play prefix of $\game$.
 Then the \emph{action decision sequence} of $\pi$ is: \[\actions(\pi) \defis j_0 \langle a_0,x_0 \rangle j_1 \langle a_1, x_1 \rangle j_2 \dots \langle a_{k-1} x_{k-1} \rangle j_k\]
 Let $\game_\A, \game_\B$ be two game arenas corresponding to an input-enabled SA $\A$, and an SA $\B$, respectively.
 Let $\sigma_1^A \in \Sigma_1(\game_\A)$, $\sigma_1^B \in \Sigma_1(\game_\B)$ be two player 1 strategies in these games.
 Strategy $\sigma_1^A$ \emph{cheats on} $\sigma_1^B$ if: 
 \[\exists \pi \in \playpref(\game_\A), \forall \pi' \in \playpref(\game_\B): \actions(\pi) = \actions(\pi') \wedge \sigma_1^B(\pi') \neq \theta \implies \sigma_1^A(\pi) = \theta \]
\end{definition}

\begin{definition}
 \label{def:alttrincl}

 Let $\game_A$ and $\game_B$ be game arenas corresponding to an input-enabled SA $\A$, and an SA $\B$, respectively.
 We say that $\game_A$ is alternating trace included in $\game_B$, denoted  $\game_A \alttraceincl  \game_B$ iff 
\begin{align*}
\forall  \sigma_2^A \in \Sigma_2(\game_A),
\exists \sigma_2^B & \in \Sigma_2(\game_B),
\forall  \sigma_1^B \in \Sigma_1(\game_B),
\exists  \sigma_1^A \in \Sigma_1(\game_A):
\sigma_1^A \text { does not cheat  on } \sigma_1^B \text{, and}\\
&\{\gtraces(\pi) \mid \pi \in \pref(\Outc(\sigma_1^A,\sigma_2^A))\} \subseteq   \{\gtraces(\pi) \mid \pi \in \pref(\Outc(\sigma_1^B,\sigma_2^B))\}
\end{align*}
\end{definition}

\begin{theorem}  \label{thm:alttrincl}
Let $\A$ and $\B$ be SAs over the same label sets and assume that $\A$ is input-enabled.
Let $\game_\A$ and $\game_\B$ be their respective underlying game arenas for the nondeterministic test assumption.
Then:
\begin{align*}
\A \ioco \B \iff \game_\A\; \alttraceincl\; \game_\B
\end{align*}

\end{theorem}

The relation between game refinement and ioco has been studied before: \cite{learningIOautomata,alternatingveanes} show that, on interface automata, $\ioco$ corresponds to alternating simulation. 
\theoremref{thm:alttrincl} differs from these results in three ways: 
(1) \cite{learningIOautomata,alternatingveanes}  compare ioco and alternating simulation on interface automata. We compare ioco on SAs versus alternating trace inclusion on games,
(2) Our games consider concurrent moves by both players;  interface automata compare different transitions of the same player. 
(3) Alternating trace inclusion is linear, whereas alternating simulation is a branching time relation. 
One could however argue that simulation and trace inclusion coincide for deterministic systems (including our SAs). However, we prefer the formulation in terms of alternating traces, because we conjecture that this formulation extends to the non-deterministic case.  

\section{Conclusions and Future Work}

We have established a fundamental connection between model-based testing and 2-player concurrent games, where
specifications are game arenas, 
test cases are game strategies, 
test case derivation is strategy synthesis, and 
conformance is alternating trace inclusion. 
This connection allows the wide plethora of game synthesis techniques to be deployed to test case generation.

The game theoretic setting spawns several game theoretic questions.
While the games we propose are concurrent because both the tester and the SUT propose moves at the same time, one could argue that they are only semi-concurrent, since only one of these moves is carried out at the same time. Therefore, we believe that the test games have various properties that do not hold for concurrent games in general. We conjecture that, whereas
concurrent games require probabilistic strategies to win reachability properties,
 our games require only deterministic strategies to win these,  and we also believe that our games are determined in that case.
 
\paragraph{\bf Acknowledgements}
 We thank Ramon Janssen and Frits Vaandrager for their comments and support.
 \vspace{-4mm}
 
\paragraph{\bf Proofs} For proofs of the theorems in this paper, we refer the reader to \url{https://petravdbos.nl/}

\bibliographystyle{eptcs}
 \bibliography{lib}

\begin{thebibliography}{10}
\providecommand{\bibitemdeclare}[2]{}
\providecommand{\surnamestart}{}
\providecommand{\surnameend}{}
\providecommand{\urlprefix}{Available at }
\providecommand{\url}[1]{\texttt{#1}}
\providecommand{\href}[2]{\texttt{#2}}
\providecommand{\urlalt}[2]{\href{#1}{#2}}
\providecommand{\doi}[1]{doi:\urlalt{http://dx.doi.org/#1}{#1}}
\providecommand{\bibinfo}[2]{#2}

\bibitemdeclare{inproceedings}{learningIOautomata}
\bibitem{learningIOautomata}
\bibinfo{author}{Fides \surnamestart Aarts\surnameend} \&
  \bibinfo{author}{Frits \surnamestart Vaandrager\surnameend}
  (\bibinfo{year}{2010}): \emph{\bibinfo{title}{Learning {I}/{O} Automata}}.
\newblock In: {\sl \bibinfo{booktitle}{International Conference on Concurrency
  Theory}}, \bibinfo{publisher}{Springer}, pp. \bibinfo{pages}{71--85},
  \doi{10.1007/978-3-642-15375-4\_6}.

\bibitemdeclare{inproceedings}{alternatingincl}
\bibitem{alternatingincl}
\bibinfo{author}{Rajeev \surnamestart Alur\surnameend}, \bibinfo{author}{Thomas
  \surnamestart Henzinger\surnameend}, \bibinfo{author}{Orna \surnamestart
  Kupferman\surnameend} \& \bibinfo{author}{Moshe \surnamestart
  Vardi\surnameend} (\bibinfo{year}{1998}): \emph{\bibinfo{title}{Alternating
  Refinement Relations}}.
\newblock In: {\sl \bibinfo{booktitle}{International Conference on Concurrency
  Theory}}, \bibinfo{organization}{Springer}, pp. \bibinfo{pages}{163--178},
  \doi{10.1007/BFb0055622}.

\bibitemdeclare{inproceedings}{bloem2016synthesizing}
\bibitem{bloem2016synthesizing}
\bibinfo{author}{Roderick \surnamestart Bloem\surnameend},
  \bibinfo{author}{Robert \surnamestart K{\"o}nighofer\surnameend},
  \bibinfo{author}{Ingo \surnamestart Pill\surnameend} \&
  \bibinfo{author}{Franz \surnamestart R{\"o}ck\surnameend}
  (\bibinfo{year}{2016}): \emph{\bibinfo{title}{Synthesizing Adaptive Test
  Strategies from Temporal Logic Specifications}}.
\newblock In: {\sl \bibinfo{booktitle}{Formal Methods in Computer-Aided
  Design}}, \bibinfo{organization}{IEEE}, pp. \bibinfo{pages}{17--24},
  \doi{10.1109/FMCAD.2016.7886656}.

\bibitemdeclare{inproceedings}{ncompletepaper}
\bibitem{ncompletepaper}
\bibinfo{author}{Petra \surnamestart {\VAN{Bos}{van den}{Van
  den}}~Bos\surnameend}, \bibinfo{author}{Ramon \surnamestart
  Janssen\surnameend} \& \bibinfo{author}{Joshua \surnamestart
  Moerman\surnameend} (\bibinfo{year}{2017}):
  \emph{\bibinfo{title}{n-{C}omplete Test Suites for {IOCO}}}.
\newblock In: {\sl \bibinfo{booktitle}{IFIP International Conference on Testing
  Software and Systems}}, \bibinfo{organization}{Springer}, pp.
  \bibinfo{pages}{91--107}, \doi{10.1007/978-3-319-67549-7\_6}.

\bibitemdeclare{inproceedings}{BrandanBB04}
\bibitem{BrandanBB04}
\bibinfo{author}{Laura~Brand\'an \surnamestart Briones\surnameend} \&
  \bibinfo{author}{Hendrik \surnamestart Brinksma\surnameend}
  (\bibinfo{year}{2004}): \emph{\bibinfo{title}{A Test Teneration Framework for
  Quiescent Real-Time Systems}}.
\newblock In: {\sl \bibinfo{booktitle}{Proc. Formal Approaches to Testing of
  Software (4th International Workshop)}}, pp. \bibinfo{pages}{71 -- 85},
  \doi{10.1007/978-3-540-31848-4\_5}.

\bibitemdeclare{inproceedings}{coveragecomplexity}
\bibitem{coveragecomplexity}
\bibinfo{author}{Krishnendu \surnamestart Chatterjee\surnameend},
  \bibinfo{author}{Luca \surnamestart De~Alfaro\surnameend} \&
  \bibinfo{author}{Rupak \surnamestart Majumdar\surnameend}
  (\bibinfo{year}{2008}): \emph{\bibinfo{title}{The Complexity of Coverage}}.
\newblock In: {\sl \bibinfo{booktitle}{Asian Symposium on Programming Languages
  and Systems}}, \bibinfo{organization}{Springer}, pp.
  \bibinfo{pages}{91--106}, \doi{10.1007/978-3-540-89330-1\_7}.

\bibitemdeclare{inproceedings}{synthesissafety}
\bibitem{synthesissafety}
\bibinfo{author}{Eric \surnamestart Dallal\surnameend}, \bibinfo{author}{Daniel
  \surnamestart Neider\surnameend} \& \bibinfo{author}{Paulo \surnamestart
  Tabuada\surnameend} (\bibinfo{year}{2016}): \emph{\bibinfo{title}{Synthesis
  of Safety Controllers Robust to Unmodeled Intermittent Disturbances}}.
\newblock In: {\sl \bibinfo{booktitle}{Decision and Control (CDC), 2016 IEEE
  55th Conference on}}, \bibinfo{organization}{IEEE}, pp.
  \bibinfo{pages}{7425--7430}, \doi{10.1109/CDC.2016.7799416}.

\bibitemdeclare{article}{david2008cooperative}
\bibitem{david2008cooperative}
\bibinfo{author}{Alexandre \surnamestart David\surnameend},
  \bibinfo{author}{Kim \surnamestart Larsen\surnameend},
  \bibinfo{author}{Shuhao \surnamestart Li\surnameend} \&
  \bibinfo{author}{Brian \surnamestart Nielsen\surnameend}
  (\bibinfo{year}{2008}): \emph{\bibinfo{title}{Cooperative Testing of Timed
  Systems}}.
\newblock {\sl \bibinfo{journal}{Electronic Notes in Theoretical Computer
  Science}}, pp. \bibinfo{pages}{79--92}, \doi{10.1016/j.entcs.2008.11.007}.

\bibitemdeclare{inproceedings}{realtimegames}
\bibitem{realtimegames}
\bibinfo{author}{Alexandre \surnamestart David\surnameend},
  \bibinfo{author}{Kim \surnamestart Larsen\surnameend},
  \bibinfo{author}{Shuhao \surnamestart Li\surnameend} \&
  \bibinfo{author}{Brian \surnamestart Nielsen\surnameend}
  (\bibinfo{year}{2008}): \emph{\bibinfo{title}{A Game-Theoretic Approach to
  Real-Time System Testing}}.
\newblock In: {\sl \bibinfo{booktitle}{Design, Automation and Test in Europe}},
  \bibinfo{organization}{IEEE}, pp. \bibinfo{pages}{486--491},
  \doi{10.1145/1403375.1403491}.

\bibitemdeclare{inproceedings}{david2009timed}
\bibitem{david2009timed}
\bibinfo{author}{Alexandre \surnamestart David\surnameend},
  \bibinfo{author}{Kim \surnamestart Larsen\surnameend},
  \bibinfo{author}{Shuhao \surnamestart Li\surnameend} \&
  \bibinfo{author}{Brian \surnamestart Nielsen\surnameend}
  (\bibinfo{year}{2009}): \emph{\bibinfo{title}{Timed Testing under Partial
  Observability}}.
\newblock In: {\sl \bibinfo{booktitle}{International Conference on Software
  Testing Verification and Validation}}, \bibinfo{organization}{IEEE}, pp.
  \bibinfo{pages}{61--70}, \doi{10.1109/ICST.2009.38}.

\bibitemdeclare{inproceedings}{outputeager}
\bibitem{outputeager}
\bibinfo{author}{Niklas \surnamestart Krafczyk\surnameend} \&
  \bibinfo{author}{Jan \surnamestart Peleska\surnameend}
  (\bibinfo{year}{2017}): \emph{\bibinfo{title}{Effective Infinite-State Model
  Checking by Input Equivalence Class Partitioning}}.
\newblock In: {\sl \bibinfo{booktitle}{IFIP International Conference on Testing
  Software and Systems}}, \bibinfo{organization}{Springer}, pp.
  \bibinfo{pages}{38--53}, \doi{10.1007/978-3-319-67549-7\_3}.

\bibitemdeclare{article}{VCGST04}
\bibitem{VCGST04}
\bibinfo{author}{Lev \surnamestart Nachmanson\surnameend},
  \bibinfo{author}{Margus \surnamestart Veanes\surnameend},
  \bibinfo{author}{Wolfram \surnamestart Schulte\surnameend},
  \bibinfo{author}{Nikolai \surnamestart Tillmann\surnameend} \&
  \bibinfo{author}{Wolfgang \surnamestart Grieskamp\surnameend}
  (\bibinfo{year}{2004}): \emph{\bibinfo{title}{Optimal Strategies for Testing
  Nondeterministic Systems}}.
\newblock {\sl \bibinfo{journal}{ACM SIGSOFT International Symposium on
  Software Testing and Analysis}}, pp. \bibinfo{pages}{55--64},
  \doi{10.1145/1013886.1007520}.

\bibitemdeclare{inproceedings}{Papadimitriou_2001}
\bibitem{Papadimitriou_2001}
\bibinfo{author}{Christos \surnamestart Papadimitriou\surnameend}
  (\bibinfo{year}{2001}): \emph{\bibinfo{title}{Algorithms, Games, and the
  {I}nternet}}.
\newblock In: {\sl \bibinfo{booktitle}{Proceedings of the thirty-third annual
  ACM symposium on Theory of computing}}, \bibinfo{publisher}{ACM Press}, pp.
  \bibinfo{pages}{749--753}, \doi{10.1145/380752.380883}.

\bibitemdeclare{inproceedings}{iotsioco}
\bibitem{iotsioco}
\bibinfo{author}{Adenilso \surnamestart Simao\surnameend} \&
  \bibinfo{author}{Alexandre \surnamestart Petrenko\surnameend}
  (\bibinfo{year}{2014}): \emph{\bibinfo{title}{Generating Complete and Finite
  Test Suite for ioco: Is It Possible?}}
\newblock In: {\sl \bibinfo{booktitle}{Proceedings of the Ninth Workshop on
  Model-Based Testing}}, pp. \bibinfo{pages}{56--70},
  \doi{10.4204/EPTCS.141.5}.

\bibitemdeclare{inproceedings}{STS13}
\bibitem{STS13}
\bibinfo{author}{Willem \surnamestart Stokkink\surnameend},
  \bibinfo{author}{Mark \surnamestart Timmer\surnameend} \&
  \bibinfo{author}{Mari\"elle \surnamestart Stoelinga\surnameend}
  (\bibinfo{year}{2013}): \emph{\bibinfo{title}{Divergent Quiescent Transition
  Systems}}.
\newblock In: {\sl \bibinfo{booktitle}{Proceedings seventh conference on Tests
  and Proofs}}, \bibinfo{series}{LNCS}, \doi{10.1007/978-3-642-38916-0\_13}.

\bibitemdeclare{incollection}{TBS11}
\bibitem{TBS11}
\bibinfo{author}{Mark \surnamestart Timmer\surnameend},
  \bibinfo{author}{Hendrik \surnamestart Brinksma\surnameend} \&
  \bibinfo{author}{Mari\"elle \surnamestart Stoelinga\surnameend}
  (\bibinfo{year}{2011}): \emph{\bibinfo{title}{Model-Based Testing}}.
\newblock In: {\sl \bibinfo{booktitle}{Software and Systems Safety:
  Specification and Verification}}, \bibinfo{series}{NATO Science for Peace and
  Security}, \bibinfo{publisher}{IOS Press}, pp. \bibinfo{pages}{1--32},
  \doi{10.3233/978-1-60750-711-6-1}.

\bibitemdeclare{incollection}{tretmans}
\bibitem{tretmans}
\bibinfo{author}{Jan \surnamestart Tretmans\surnameend} (\bibinfo{year}{2008}):
  \emph{\bibinfo{title}{Model {B}ased {T}esting with {L}abelled {T}ransition
  {S}ystems}}.
\newblock In: {\sl \bibinfo{booktitle}{Formal methods and testing}},
  \bibinfo{publisher}{Springer}, pp. \bibinfo{pages}{1--38},
  \doi{10.1007/978-3-540-78917-8\_1}.

\bibitemdeclare{inproceedings}{alternatingveanes}
\bibitem{alternatingveanes}
\bibinfo{author}{Margus \surnamestart Veanes\surnameend} \&
  \bibinfo{author}{Nikolaj \surnamestart Bj{\o}rner\surnameend}
  (\bibinfo{year}{2010}): \emph{\bibinfo{title}{Alternating Simulation and
  {IOCO}}}.
\newblock In: {\sl \bibinfo{booktitle}{IFIP International Conference on Testing
  Software and Systems}}, \bibinfo{organization}{Springer}, pp.
  \bibinfo{pages}{47--62}, \doi{10.1007/978-3-642-16573-3\_5}.

\bibitemdeclare{article}{confrelandtestobj}
\bibitem{confrelandtestobj}
\bibinfo{author}{Ren{\'e} \surnamestart de~Vries\surnameend} \&
  \bibinfo{author}{Jan \surnamestart Tretmans\surnameend}
  (\bibinfo{year}{2001}): \emph{\bibinfo{title}{Towards Formal Test Purposes}}.
\newblock {\sl \bibinfo{journal}{Formal Approaches to Testing of Software,
  FATES'01: A Satellite Workshop of CONCUR'01 Proceedings}}, pp.
  \bibinfo{pages}{61--76}.

\bibitemdeclare{article}{nodecoverage}
\bibitem{nodecoverage}
\bibinfo{author}{Farn \surnamestart Wang\surnameend}, \bibinfo{author}{Sven
  \surnamestart Schewe\surnameend} \& \bibinfo{author}{Jung-Hsuan \surnamestart
  Wu\surnameend} (\bibinfo{year}{2015}): \emph{\bibinfo{title}{Complexity of
  Node Coverage Games}}.
\newblock {\sl \bibinfo{journal}{Theoretical Computer Science}}, pp.
  \bibinfo{pages}{45--60}, \doi{10.1016/j.tcs.2015.02.002}.

\end{thebibliography}
 
\end{document}